\documentclass[showpacs,amsmath,superscriptaddress,pre]{revtex4}
\usepackage{epsfig}
\usepackage{latexsym}
\usepackage{wasysym}

\begin{document}

\preprint{EV/25-07-2008}

\title{Work distributions in the $T=0$ Random Field Ising Model}

\author{Xavier Illa}
\affiliation{Department of Applied Physics, Helsinki University of 
Technology,  P.O.Box 1100, FIN-02015 HUT, Finland}
\author{Josep Maria Huguet}
\affiliation{Departament de F\'{\i}sica Fonamental, Universitat de Barcelona 
\\ Mart\'{\i} i Franqu\`es 1, Facultat de F\'{\i}sica, 08028 Barcelona, 
Catalonia}
\author{Eduard Vives}
%
%
\affiliation{ Departament d'Estructura i Constituents de la Mat\`eria,
  Universitat de  Barcelona \\ Mart\'{\i} i Franqu\`es 1 ,  
  Facultat de F\'{\i}sica,
  08028 Barcelona, Catalonia}

\date{\today}

\begin{abstract}
  We perform  a numerical study of the  three-dimensional Random Field
  Ising Model at $T=0$. We compare work distributions along metastable
  trajectories obtained  with the  single-spin flip dynamics  with the
  distribution  of  the   internal  energy  change  along  equilibrium
  trajectories.   The  goal  is  to  investigate  the  possibility  of
  extending the  Crooks fluctuation theorem  \cite{Crooks1999} to zero
  temperature when,  instead of the standard  ensemble statistics, one
  considers the ensemble generated  by the quenched disorder.  We show
  that  a simple  extension of  Crooks fails  close to  the disordered
  induced equilibrium phase  transition due to the fact  that work and
  internal energy distributions are very asymmetric.
\end{abstract}


\maketitle

\section{Introduction}

The $T=0$ Random Field Ising Model  is a prototype model for the study
of collective  phenomena in disordered systems.   Although it neglects
thermal fluctuations,  it contains essential  competitions between the
quenched  disorder,  the ferromagnetic  interaction  and the  external
applied  field.   The  model  can  be  numerically  studied  from  two
different points  of view:  on the one  hand, the exact  ground state
calculation \cite{Hartmann2001,Middleton2002,Dukovski2003,Wu2005}
provides   an   approach  to   the
equilibrium phase diagram.  On the other the use of a local relaxation
dynamics based on single spin-flips  provides a good framework for the
understanding     of     avalanche     dynamics     and     hysteresis
\cite{Sethna1993,Sethna2005},   which   is   closer  to   experimental
observations.  In this sense,  the model  is a  good workbench  for the
comparison of equilibrium and out-of-equilibrium trajectories.

A  number  of  non-equilibrium  work theorems  \cite{Sevick2008}  have
received a lot  of attention in the last  10 years, particularly after
the  work  of  Jarzynski in  1997  \cite{Jarzynski1997,Jarzynski2006}.
These  theorems relate  in  different ways  the  distribution of  work
performed  on a  system which  is driven  out of  equilibrium  to some
equilibrium  thermodynamic  properties. One  example  is the  original
Jarzinsky's equality: $\langle e^{-\beta W} \rangle = e^{-\beta \Delta
F}$,  where $W$  is the  work performed  on the  system that  has been
driven (out-of-equilibrium)  by varying an  external control parameter
$H$ changing from $H(0)$ to $H(1)$,  and $\Delta F$ is the free energy
difference  between two  states $0$  and  $1$ that  correspond to  the
equilibrium states  at $H(0)$ and $H(1)$.  The  average $\langle \cdot
\rangle$ should  be understood as  obtained after many  repetitions of
the driving  process. The system  is assumed to  be in contact  with a
heat bath  at temperature $\beta^{-1}$ which  generates an statistical
ensemble  of  copies  of  the  system  which is  the  source  of  work
fluctuations.   Most  of   these  theorems  have  been  experimentally
verified. \cite{Collin2005,Trepagnier2004}

The goal of this paper is to investigate how such kind of theorems can
be  extended to systems  at  $T=0$.   In  such a  case,  although
equilibrium  states  and   out-of-equilibrium  trajectories  are  well
defined, there  are no  thermal fluctuations. A  priori it  seems that
there is no statistical ensemble over which one can define probability
distributions or averages. The idea we want to test is whether or not,
for  systems  with quenched  disorder,  the  thermal  ensemble can  be
substituted by  the ensemble of different realizations  of disorder and
still some work theorems can be applied.

In order to  perform the $T\rightarrow 0$ limit  we choose a different
but  related  work  theorem   which  is  Crooks  fluctuations  theorem
\cite{Crooks1999,Crooks2000}.  The advantage will be that it allows to
derive an  equality that can  be extrapolated to the  $T\rightarrow 0$
limit.  This theorem can be written as: 
\begin{equation}
\frac{P_F(W)}{P_R(-W)} = e^{\beta(W-\Delta F)}
\label{Crooks}
\end{equation}
In  this case  $P_F(W)$ and  $P_R(-W)$ are  the  probability densities
corresponding  to the  out-of-equilibrium work  performed on  a system
driven  forward from  $H(0)$ to  $H(1)$ and  reversely from  $H(1)$ to
$H(0)$. The Crooks fluctuation theorem has been formulated 
under several assumptions: the driven systems must be finite,
classical and coupled to a thermal bath. Moreover, the dynamics should be stochastic, Markovian and reversible and the entropy production should be odd under time reversal. 
Some of these assumptions are clearly not accomplished at the work at 
hand. For instance the thermal bath is at $T=0$ and the dynamics is deterministic. 
Nevertheless we have been a little speculative and investigated the possibility 
of extending the Crooks fluctuation theorem to $T=0$ for systems with 
quenched disorder.
 
From Eq.~(\ref{Crooks}) one can derive  that the value $W^*$ for which
the two probability densities  are equal, i.e.  $P_F(W^*) = P_R(-W^*)$
satisfies $W^* =  \Delta F$.  This result is  particularly suitable to
be investigated at $T \rightarrow 0$. Note that in the low temperature
limit,  the  only  possibility  to   have  a  crossing  point  of  the
probability densities at  $W^*$ is that $W^* =  \Delta F(T \rightarrow
0) = \Delta  U$, so that the  diverging behaviour of  $\beta$ might be
cancelled to get a finite limit $P_F(W^*)/P_R(-W^*) \rightarrow 1$.

\section{Model}

For our investigation we consider the Random Field Ising Model. A set
of $N$ spin variables $S_i=\pm 1$  with $i=1,\cdots N$ is defined on a
regular 3D cubic lattice with linear size $L$ ($N=L \times L \times L$).
We are  interested in  following the response  of the  order parameter (magnetization) 
$M=\sum_{i=1}^{N} S_i$ as a  function of the external driving
field $H$. The Hamiltonian (magnetic enthalpy) of the model is defined
as:
\begin{equation}
\label{hamiltonian}
{\cal  H}=U-HM 
\end{equation}
\begin{equation}
U= -   \sum_{ij}  S_i  S_j   -  \sum_{i=1}^N  h_i   S_i  
\label{energy}
\end{equation}
%
%
where the first sum in Eq.~\ref{energy},  extending over nearest neigbours pairs, accounts
for a ferromagnetic exchange interaction, the second term in Eq.~\ref{energy}
includes the interaction with the quenched disorder  and the second term in Eq.~\ref{hamiltonian} accounts for the interaction with  the external field $H$ that will  be used as the
driving  parameter. 
The  local  fields $h_i$  are independent  random variables, quenched, Gaussian distributed 
with zero mean and standard deviation $\sigma$.
All along the paper we will use small letters ($u=\frac{U}{N}$,$m=\frac{M}{N}\ldots$) 
to refer to intensive magnitudes.

The equilibrium properties of this model can be studied at $T=0$.  For
a given realization  of the random fields $\{h_i\}$  and a given value
of  $H$, the  exact ground  state can  be found  in a  polynomial time
\cite{Middleton2002b}. Moreover optimized  algorithms can also be used
to  obtain the  full sequence  of ground-sates  as a  function  of $H$
between  the  two saturated  states  $\{S_i=-1\}$  at $H=-\infty$  and
$\{S_i=1\}$  at $H=\infty$  \cite{Frontera2000}. In  the thermodynamic
limit the  3D model exhibits a  first order phase  transition at $H=0$
and for  $\sigma < \sigma_c^{eq}=2.27$ \cite{Hartmann2001,Middleton2002}. 
Although finite  systems do not  exhibit a  true phase  transition, there  is a
region of $\sigma$ and $H$ where the correlation length is of the same
order  as the  system  size.   This leads  to  collective effects  and
correlations involving all the spins of the system. 

Concerning  the  non-equilibrium  trajectories,  the  model  has  been
intensively studied by using the so called single-spin flip metastable
dynamics  at $T=0$  which can  be understood  as a  $T  \rightarrow 0$
Glauber dynamics.  It consists  in adiabatically sweeping the external
field and  relaxing individual spins according to  their local energy,
i.e. spins should align with the local field
\begin{equation}
F_i = H + h_i + \sum_k S_k
\end{equation}
where the sum  extends over the neigbours of spin  $i$.  Note that the
reversal of  a spin at  a given field  $H$ may induce the  reversal of
some of its  neigbours at the same field  $H$.  Such collective events
are the so  called avalanches that end when all  the spins are stable.
Only after the avalanches have  finished, the external field is varied
again.  It  has been  shown that, during  a monotonous driving  of the
external  field, no reverse  spin-flips may  occur and,  moreover, the
model exhibits the so called  abelian property: i.e.  the final states
do not  depend on the order  in which the unstable  spins are flipped.
In the  thermodynamic limit, the system with  this metastable dynamics
also    shows    a    critical    point    at    $\sigma_c^{met}=2.21$
\cite{Sethna1993, PerezReche2003}, below which  the metastable trajectory exhibits
a discontinuity. 

Note   that  avalanches   are   the  source   of  energy   dissipation
\cite{Ortin1998,Illa2005}.   If  we think  about  an increasing  field
trajectory, the triggering  spin of each avalanche flips  at the field
$H$ that corresponds to  $F_i=0$. This spin, therefore, flips without
energy loss $\Delta {\cal H} = \Delta U -H \Delta  M=0$.  But the subsequent
unstable spins flip (at a fixed external field $H$) 
with an  external force $F_i>0$ giving raise to an
energy  loss (associated to each individual spin flip) 
$Q=-2F_i<0$  with $Q=\Delta  U -W$.   For  the decreasing
field branches, $F_i<0$  but the energy loss is  then $Q=2F_i<0$ since
the spins flip from $1$ to $-1$.
 
\begin{figure}[ht]
\begin{center}
  \epsfig{file=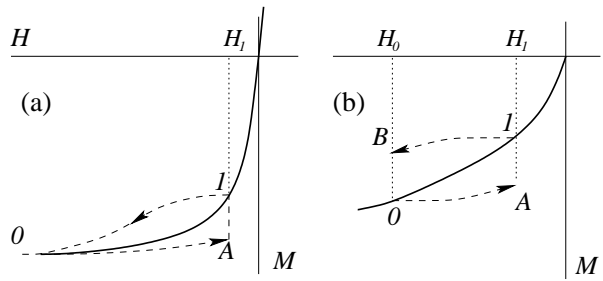,width=7.9cm,clip=}
\end{center}
\caption{\label{FIG1}  Schematic representation  of the  two different
strategies that have been used.  (a) corresponds to the first strategy
in  which  the  starting  point   is  the  saturated  state $0$ ($m=-1$,$H=-\infty$),
and  (b) corresponds  to the  second  strategy, starting  from the  equilibrium
state $0$ at  a finite  field $H_0$.   Dashed  lines  correspond to  metastable
trajectories and the continuum line indicates the equilibrium states.}
\end{figure}

Note that in  this discussion we are using  the standard definition of
work that is not a state function. It is computed as a sum over the 
($k=1,2\ldots$) spins that flip along the trajectory
\begin{equation}
\label{work}
W=\sum_k H_k \Delta M_k,
\end{equation}
where $\Delta M_k=2$. Recently    there    has   been    a    discussion
\cite{Vilar2008,Imparato2007,Horowitz2008,Peliti2008}  on  which  is  the
most suitable  definition of work  to be used in  such non-equilibrium
work   theorems.    Already   from   the   initial   Jarzynski   works
\cite{Jarzynski1997}, it was proposed that  if the system is driven by
controlling the  external field, the convenient definition  of work is
the integral  over the  trajectory $V=\int M  dH$.  Note that  along a
metastable trajectory  , the  two definitions of  work are  related by
$W=\Delta  (HM)-V$.   Without  going  into the  discussion,  we  will
numerically test here the two  possibilities. In Sec.  \ref{W} we will
test whether the  crossing point $W^*$ of the  histograms $P_F(W)$ and
$P_R(-W)$ satisfies
\begin{equation}
W^* = \langle \Delta U \rangle
\end{equation}
and, in Sec. \ref{V} we will  test whether the crossing point $V^*$ of
the  histograms  $P_F(V)$ and  $P_R(-V)$  satisfies the  corresponding
equation:
\begin{equation}
V^* = \langle \Delta U - \Delta(HM) \rangle
\end{equation}
The  test  of  the  two  hypothesis will  require  slightly  different
strategies (as indicated in Fig. \ref{FIG1}) since $V$ can not be 
computed for trajectories starting at saturation ($H\rightarrow \pm \infty$). 
In  the  first  strategy  (Sec.   \ref{W}), we  will  compute  forward
trajectories starting from the state $0$, saturated with $m=-1$ at
$H=-\infty$ and sweep the  field adiabatically from $-\infty$ to $H_1$
until the metastable  state $A$ is reached.  Then  we will compute the
equilibrium state $1$ corresponding to the field $H_1$ and perform the
reverse trajectory from  $1$ to $0$ with the  metastable dynamics.  We
will  compute $\Delta u =(U_1-U_0)/N$ and  the following
works (using the 'standard' definition):
\begin{eqnarray}
\label{w01}
w_{0 \rightarrow 1}&=&\frac{W_{0 \rightarrow 1}}{N} =  \frac{1}{N} \int_0^A  H dM + \frac{1}{N} \int_A^1 H dM =
 \frac{1}{N} \int_0^A  H dM + \frac{1}{N} H (M_1-M_A)
\\  w_{1 \rightarrow 0} &=& \frac{W_{1 \rightarrow 0}}{N} = \frac{1}{N} \int_1^0 H dM
\end{eqnarray}
where  $M_A$  and $M_1$  are  the  magnetizations  of states  $A$  and
$1$. The integrals  are computed along the trajectories schematically represented in 
Fig.~\ref{FIG1}(a) using Eq.~\ref{work}. 
Note also that the second integral in Eq.~\ref{w01} can be computed 
since the process $A \rightarrow 1$ takes place at constant field $H_1$. 

In the second  strategy (Sec. \ref{V}), we will  start from a computed
equilibrium state at $H_0$,  perform a metastable trajectory until the
state $A$ is  reached at $H_1$.  Then we  will compute the equilibrium
state $1$  at $H_1$ and  perform a metastable trajectory  driving back
until reaching the  state $B$ at $H_0$.  In this case  we will use the
alternative definition of work: 
\begin{eqnarray}
v_{0 \rightarrow 1}=v_{0 \rightarrow A}=\frac{V_{0 \rightarrow 1}}{N}  &=&  \frac  {V_{0 \rightarrow A}}{N} =  \frac{1}{N}
\int_0^A M dH, \\ v_{1 \rightarrow 0}=v_{1 \rightarrow B}=\frac{V_{1 \rightarrow 0}}{N} &=& \frac{V_{1 \rightarrow B}}{N}=
\frac{1}{N} \int_1^B M dH.
\end{eqnarray}
The integrals are computed along the trajectories schematically represented in 
Fig.~\ref{FIG1}(b) using $ V=\sum_k M_k \Delta H_k $ where $k$ accounts for 
the sequence of interavalanche field increments
$\Delta H_k$  occuring at constant $M_k$ along the trajectory.

.
\begin{figure}[ht]
\begin{center}
  \epsfig{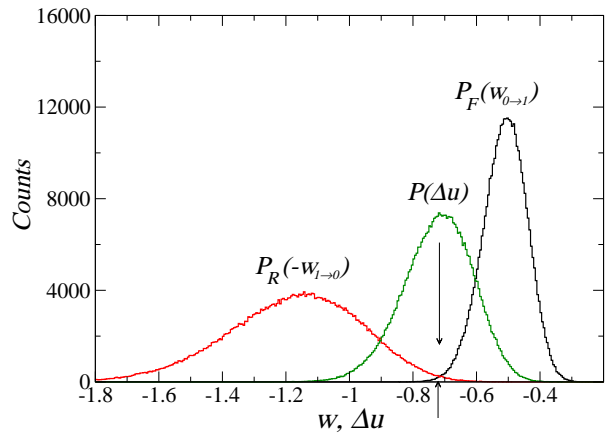}
\end{center}
\caption{\label{FIG2}   (Color   on   line)  Example   of   histograms
corresponding to $P_F(w_{0 \rightarrow 1})$, $P_R(-w_{1 \rightarrow 0})$ and $P(\Delta u)$.  The
arrow  pointing downwards  indicates the  position of  the  maximum of
$P(\Delta  u)$  and  the  arrow  pointing upwards  the  average  value
$\langle \Delta u \rangle $.  The example corresponds to a system with
$\sigma=4$, $H_1=-0.35$  and $L=12$. Histograms have  been obtained by
cumulating  data  corresponding  to  $4\times  10^5$  realizations  of
disorder. }
\end{figure} 
%

\section{First strategy}
\label{W}

Fig.   2  shows  histograms  corresponding  to  the  distributions  of
$w_{0 \rightarrow 1}$, $\Delta  u$ and $-w_{1 \rightarrow 0}$ for $\sigma=4$,  $H_1=-0.35$ and a
system size $L=12$. The computed  histograms give an estimation of the
corresponding probability  densities $P_F(w_{0 \rightarrow 1})$, $P_R(-w_{1 \rightarrow 0})$ and
$P(\Delta  u)$. Four important  points should  be realized  from this
example: (i) For this value of $\sigma$ and field $H_1$, the densities
look symmetric  and Gaussian. This will  not be the case  when $H_1$ and
$\sigma$ are close  to the disorder  induced phase  transition at
$\sigma_c^{eq}$ and  $H_1=0$.  (ii) Second, the  distribution of $\Delta
u$ has a width similar to those of the out-of-equilibrium works.  This
is different from  what happens in the analysis  of Crooks fluctuation
theorem at  finite $T$.  Typically,  equilibrium thermal fluctuations
are much smaller than work  fluctuations. (iii) For increasing systems size
the histograms become narrower and then the crossing points are harder to locate
(iv) Finally, note that for this case the  hypothesis that we are testing  
is fulfilled: the peak in $P(\Delta  u)$ as  well as the  average $\langle \Delta  u \rangle$
coincide  with   the  crossing   point  $w^*$  of   $P_F(w_{0 \rightarrow 1})$  and
$P_R(-w_{1 \rightarrow 0})$.
%
\begin{figure}[ht]
\begin{center}
  \epsfig{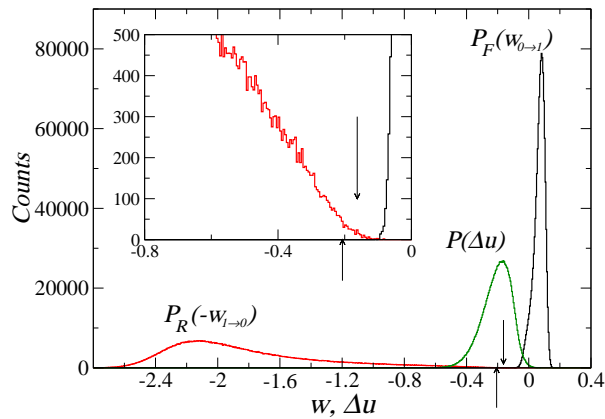}
\end{center}
\caption{\label{FIG3}   (Color   on   line)  Example   of   histograms
corresponding to $P_F(w_{0 \rightarrow 1})$, $P_R(-w_{1 \rightarrow 0})$ and $P(\Delta u)$.  The
arrow  pointing downwards  indicates the  position of  the  maximum of
$P(\Delta  u)$  and  the  arrow  pointing upwards  the  average  value
$\langle \Delta u \rangle $.  The example corresponds to a system with
$\sigma=3$, $H_1=0.1$ and $L=12$. The histograms have been computed by
cumulating  data  corresponding to  $1.6\times  10^6$ realizations  of
disorder. The inset shows a detailed  view of the crossing point of the
$P_F(w_{0 \rightarrow 1})$ and $P_R(-w_{1 \rightarrow 0})$ histograms.}
\end{figure} 
%
%
\begin{figure}[ht]
\begin{center}
  \epsfig{file=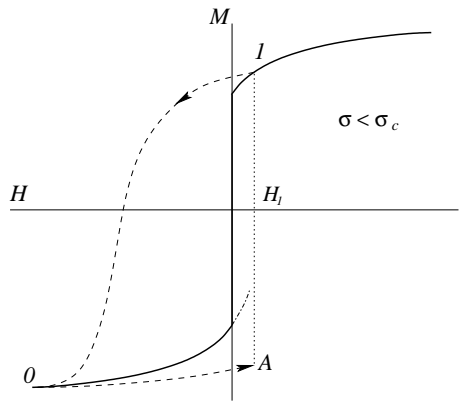,width=6.0cm,clip=}
\end{center}
\caption{\label{FIG4}  Schematic representation  of  the case  $\sigma
\lesssim  \sigma_c^{eq}$  and  $H_1>0$.   Dashed lines  correspond  to
metastable trajectories, continuous lines to the equilibrium trajectory
and the  dashed-dotted line  to the non  typical equilibrium states  that are
responsible   for   the    high   asymmetry   of   the   distribution
$P_R(-w_{1 \rightarrow 0})$.}
\end{figure}
%
Fig.~\ref{FIG3}   shows  histograms   corresponding   to  $P(w_{0 \rightarrow 1})$,
$P(\Delta u)$ and $P(-w_{1 \rightarrow 0})$  for $\sigma=3$, $H_1=0.1$ and a system
size $L=12$.  Note  than in this case, the  work distributions as well
as  the energy  distribution  are very  asymmetric.   This is  because
although in this  case we expect $M_1>0$ since  $H_1>0$, a certain non
vanishing  fraction of the  equilibrium states  still has  $M_1<0$, as
schematically  indicated  in  Fig.\ref{FIG4}.   In  other  words,  for
certain  realizations of  disorder,  the  state $1$  turns  out to  be
non-typical  and  previous   to  the equilibrium transition from $M<0$ to $M>0$.
Consequently,  the distribution of  reverse works $w_{1 \rightarrow 0}$  widens enormously. As  can be seen  the proposed equality  $w^* =  \Delta u$  clearly fails  in this
case. The  relative error is 
 $| w^* -  \langle \Delta u \rangle|  / | \langle \Delta u \rangle| \sim 0.25$. 

\begin{figure}[ht]
\begin{center}
  \epsfig{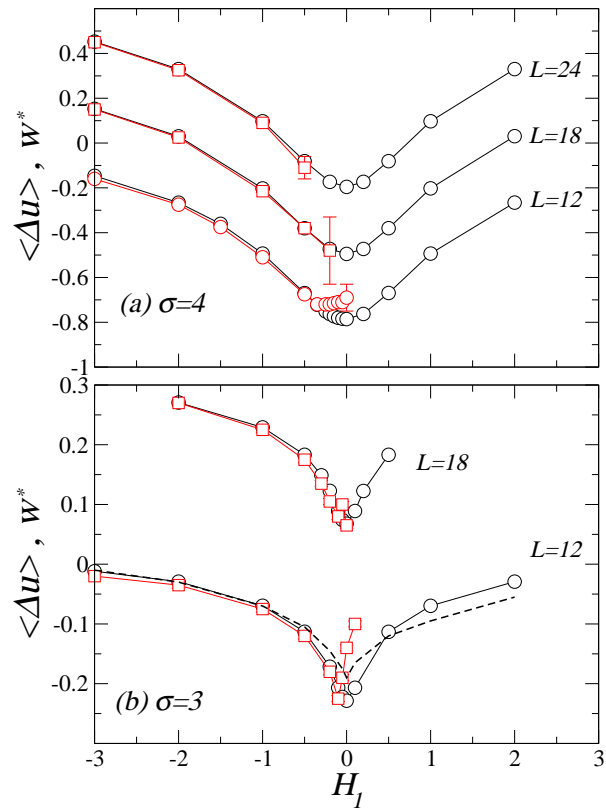}
\end{center}
\caption{\label{FIG5} (Color on line) Comparison of the crossing point
of the histograms $w^*$ ($\Box$) with  $\langle \Delta u \rangle$ 
($\Circle$) as a function of  $H_1$,  corresponding  to  (a)  $\sigma=4$ and  (b)  $\sigma=3$  for different  values of $L$  as indicated  by the  legend.  For  the case
$L=12$ and  $\sigma=3$ the dashed  line indicated the position  of the
maximum of $P(\Delta  u)$.  Data for $L=18$ ($L=24$)  has been shifted
0.3  (0.6) units  along  the vertical  axis  in order  to clarify  the
picture.}
\end{figure} 
%
We  have performed a  detailed study  of the  deviation of  $w^*$ from
$\langle \Delta u \rangle $ for  different values of $\sigma$, $H_1$  and $L$. Examples
of the  results are presented in  Fig. \ref{FIG5}. Note  that the data
corresponding to  the crossing  point $w^*$ exhibits  increasing error
bars  with  increasing  $H_1$.   This  is because  the  histograms  of
$P_F(w_{0 \rightarrow 1})$, $P_R(-w_{1 \rightarrow 0})$  become more and more  separated and the
finding of  the intersection requires  more and more  statistics. (This
problem  is accentuated  for larger  values  of $H_1$  and for  larger
system  sizes).   One  can   conclude  therefore  than   the  proposed
extrapolation of  Crooks theorem, given by  the equality $w^*=\langle
\Delta u \rangle$ fails in the  region of $\sigma$ and $H_1$ where the
system exhibits a collective behaviour (i.e. the correlation length is
similar  to  the  system  size)   because  of  the  proximity  of  the
equilibrium   phase   transition.    Consistently,   we   observe   in
Fig.  \ref{FIG5}(b) (comparing  the data  corresponding to  $L=12$ and
$L=18$ at  $H_1\sim 0$) that  the region of breakdown  becomes smaller
when  the system  size $L$  is  larger.  As  expected, increasing  the
system  size with  fixed correlation  length decreases  the collective
effects.

\section{Second strategy}
\label{V}

\begin{figure}[ht]
\begin{center}
  \epsfig{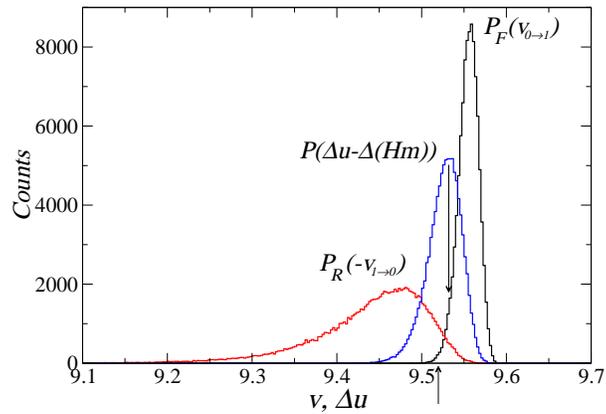}
\end{center}
\caption{\label{FIG6}   (Color   on   line)  Example   of   histograms
corresponding  to  $P_F(v_{0 \rightarrow 1})$,  
$P_R(-v_{1 \rightarrow 0})$  and $P\left ( \Delta  u - \Delta (H m) \right )$. 
Data corresponds  to $H_0=-10.0$,$H_1=-0.4$, $\sigma=3$
and $L=12$. Histograms are  computed by cumulating $10^5$ realizations
of disorder. The arrows pointing down (up) indicate the
peak (average value) of the distribution $P(\Delta u - \Delta (HM))$. }
\end{figure}

Fig. \ref{FIG6} shows histograms corresponding to the distributions of
$v_{0 \rightarrow 1}$ and $-v_{1 \rightarrow 0}$ for  the case $L=12$, $\sigma=3$, $H_1=-10$ and
$H_2=-0.4$. As  can be seen  the intersection point $v^*$  agrees very
well with the average value of $\Delta u - \Delta(Hm)$. In this case we
have also studied  if this agreement is equally  good in other regions
of the phase diagram.
%
%
\begin{figure}[ht]
\begin{center}
  \epsfig{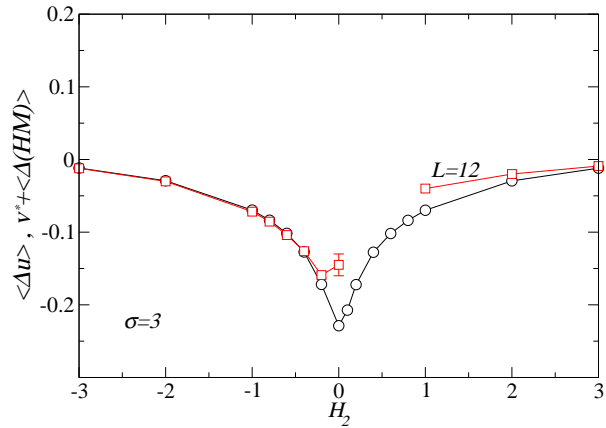}
\end{center}
\caption{\label{FIG7} (Color on line) Comparison of the crossing point
$v^*$ ($\Box$) of the histograms $P_F(v_{0 \rightarrow 1})$ and $P_R(-v_{1 \rightarrow 0})$ with $\langle
\Delta u -\Delta (Hm) \rangle$ ($\Circle$) as a function of $H_2$, corresponding to
$\sigma=3$  and  $L=12$.  Data  corresponds  to  averages over  $10^5$
realizations of disorder}
\end{figure} 
%
Fig \ref{FIG7}  shows a  comparison of $\langle \Delta  u \rangle$ and $v^*  + \langle
\Delta (H m) \rangle$. On the right  hand side, the value  of $v^*$ has
not  been obtained by  a direct  location of  the intersection  of the
histograms  but   locating the mid  point  between   the positions of the maxima of the  histograms corresponding to $P_F(v_{0 \rightarrow 1})$ and $P_R(-v_{1 \rightarrow 0})$
which are far separated but very  symmetric. 
As can  be seen, as  occurs with strategy  1, the
agreement  $v^*  \simeq  \langle \Delta  u  -  \Delta  (Hm) \rangle$  fails  when  $H_2$
approaches the phase transition region.

\section{Summary and conclusion}
\label{Conslusion}
We  have investigated  the possibility of extending some  work fluctuation
theorems to $T=0$  for systems with quenched disorder.   In this case,
thermal averages should be  substituted by disorder averages.  We have
proposed an hypothesis based on the extrapolation of Crooks theorem to
the  $T  \rightarrow 0$  limit  and  we  have numerically  tested  its
validity.  The  hypothesis is that if one  considers the distributions
of  non-equilibrium  work corresponding  to  a  forward and  backwards
trajectory,   the   crossing   point   $w^*$  of   $P_F(w_{0 \rightarrow 1})$   and
$P_R(-w_{1 \rightarrow 0})$,  is  equal  to  the  average of  the  internal  energy
difference $\Delta u=u_{1}-u_2$. The  investigation has been  done by
considering two  strategies, based on the two  possible definitions of
work that  have been discussed in  the literature: $w=\int  H dm$ and
$v=\int mdH$.

The  reported numerical   investigation  indicates   that,  for  both
strategies, the  formulated hypothesis is  valid when  the system
does not  behave collectively. Thus, far from the
equilibrium critical point, where correlations are small compared to
system size, the distribution $p(\Delta u)$ is very symmetric and the
average (equilibrium) value $\langle \Delta u \rangle$ can be obtained from the crossing
point $w^*$ of the distributions of the non equilibrium works.

When the system is close to the critical point a definitive conclusion 
can not be stated. Data for small $L$ suggest that when the correlation
length approaches the system size $L$, the distribution $p(\Delta u)$ is very 
assymetric and there is no connection between the equilibrium value $\langle \Delta u \rangle$ 
and the crossing point $w^*$ of the non-equilibrium work distributions.
A bigger computational effort (not affordable at present) would be needed in order
to compute the accurate histograms for $L>24$. 
The presented results may provide some clues for future  investigations in  order to
extend work fluctuation theorems.

\acknowledgments This  work has received financial  support from CICyT
(Spain),    project   MAT2007-61200,   CIRIT    (Catalonia),   project
2005SGR00969,    EU    Marie     Curie    RTN    MULTIMAT,    Contract
No. MRTN-CT-2004-5052226 and the Academy of Finland. X.I. acknowledges
the hospitality of ECM Department (Universitat de Barcelona) where the
work  at hand  was to  a large  degree completed.  We  also acknowledge
fruitful discussions with F. Ritort, A. Planes, and M.J. Alava.

\end{document}